\documentclass[preprint,showpacs,preprintnumbers,amsmath,amssymb,nofootinbib]{revtex4}
\input psfig.sty
\def \be {\begin{equation}}

\def \ee {\end{equation}}
\def \bea {\begin{eqnarray}}
\def \eea {\end{eqnarray}}

\usepackage{graphicx}
\usepackage{dcolumn}
\usepackage{bm}
\usepackage{epsfig}

\begin{document}

\title{The Zero Active Mass Condition in \\  Friedmann-Robertson-Walker Cosmologies}

\author{F. Melia}\thanks{John Woodruff Simpson Fellow} \email{fmelia@email.arizona.edu}

\vskip 1.5cm
\affiliation{Department of Physics, The Applied Math Program, and Department of Astronomy,
The University of Arizona, AZ 85721, USA}


\begin{abstract}
\noindent 
Many cosmological measurements today suggest that the Universe is
expanding at a constant rate. This is inferred from the observed
age versus redshift relationship and various distance indicators,
all of which point to a cosmic equation of state (EoS) $p=-\rho/3$,
where $\rho$ and $p$ are, respectively, the total energy density and
pressure of the cosmic fluid. It has recently been shown that this
result is not a coincidence and simply confirms the fact that the
symmetries in the Friedmann-Robertson-Walker (FRW) metric appear
to be viable only for a medium with zero active mass, i.e.,
$\rho+3p=0$. In their latest paper, however, Kim, Lasenby and
Hobson have provided what they believe to be a counter argument
to this conclusion. Here, we show that these authors are merely
repeating the conventional mistake of incorrectly placing the
observer {\it simultaneously} in a comoving frame, where the
lapse function $g_{tt}$ is coordinate dependent when $\rho+3p\not=0$,
and a supposedly different, free-falling frame, in which
$g_{tt}=1$, implying no time dilation. We demonstrate that the
Hubble flow is not inertial when $\rho+3p\not=0$, so the comoving
frame is generally not in free fall, even though in FRW, the comoving 
and free-falling frames are supposed to be identical at every
spacetime point. So this confusion of frames not only constitutes
an inconsistency with the fundamental tenets of general relativity
but, additionally, there is no possibility of using a gauge
transformation to select a set of coordinates for which
$g_{tt}=1$ when $\rho+3p\not=0$.
\end{abstract}

\maketitle

\section{Introduction}
The standard model of cosmology ($\Lambda$CDM) is quite successful in
accounting for many observations, in part due to its rather large number
of free parameters. These include the spatial curvature constant
$k$, the Hubble constant $H_0$ and the scaled matter ($\Omega_{\rm m}$),
radiation ($\Omega_{\rm r}$), and dark-energy ($\Omega_{\rm de}$) densities.
In terms of the critical density $\rho_{\rm c}\equiv 3c^2{H_0}^2/8\pi G$,
one writes $\Omega_i\equiv \rho_i/\rho_{\rm c}$, where $i$ represents the
energy density $\rho_{\rm m}$, $\rho_{\rm r}$, or dark energy $\rho_{\rm de}$,
as the case may be. Without invoking priors, one must typically optimize at
least five unspecified parameters, all of which can be adjusted to fit the data.

Given this wide latitude of possible outcomes for the expansion history of
the Universe, it is therefore very surprising to see that the optimization of
model parameters, especially using anisotropies in the cosmic microwave
background (CMB) \cite{Bennett2003,Spergel2003,Ade2014},
reveals a universal expansion with an average acceleration of zero (within the
measurement errors) over a Hubble 
time ${H_0}^{-1}$ \cite{Melia2003,Melia2007,MeliaAbdelqader2009,MeliaShevchuk2012,Melia2016}.
Another way to characterize this empirical result, in terms
of the total pressure $p=p_{\rm r}+ p_{\rm m}+p_{\rm de}$ and energy
density $\rho=\rho_{\rm r}+\rho_{\rm m}+\rho_{\rm de}$, is to note
that averaged over a Hubble time, the quantity $\langle p/\rho\rangle$
has the value $-1/3$ just at this moment, when we happen to be looking.
Yet the combination of $\rho_{\rm m}$, $\rho_{\rm r}$ and $\rho_{\rm de}$
could have produced a wide assortment of epochs with deceleration and acceleration.

This outcome is not merely surprising \cite{Kim2016}. In the context of $\Lambda$CDM, 
the condition $\langle p/\rho \rangle =-1/3$ can be achieved only once in the
entire (presumably infinite) history of the Universe, making it astonishingly
unlikely. There is clearly physics behind this zero active mass condition.
In a recent paper \cite{Melia2016}, we showed that the Friedmann-Robertson-Walker
(FRW) metric is in fact valid only for a cosmic fluid with an equation of
state (EoS) $\rho+3p=0$, so the condition $\langle p/\rho\rangle=-1/3$
is independent of time. In other words, no matter how imperfect or incomplete
our parametrization of the standard model happens to be, the optimized
parameter values must always yield a zero average acceleration, regardless
of when we make the measurements.

Kim et al. \cite{Kim2016} have challenged this conclusion by providing what they believe
to be a counter argument to the zero active mass condition in FRW. In this
paper, however, we demonstrate that their claim is based on the
conventional mistake of writing the metric from the perspetive of an
observer who is {\it simultaneously} in a non-inertial comoving frame, where a
time dilation is unavoidable for an accelerated expansion of the spatial
coordinates, and in a different, free-falling frame, with no measurable
acceleration, thereby ``assigning" without justification a constant value
of one to the lapse function $g_{tt}$.

\section{The Lapse Function in FRW}
With the adoption of the Cosmological principle, the spacetime metric may be
simplified to the FRW form
\begin{equation}
ds^2=c^2dt^2-a^2(t)\left[dr^2(1-kr^2)^{-1}+r^2(d\theta^2+\sin^2\theta\,d\phi^2)\right]\;,
\end{equation}
in terms of the cosmic time $t$, the comoving radius $r$, the universal expansion
factor $a(t)$, and the angular coordinates $\theta$ and $\phi$ in the comoving
frame. The spatial curvature constant $k$ takes on the values $(-1,0,+1)$,
for an open, flat, or closed universe, respectively.

Two rarely discussed issues with this metric are (1) that its
lapse function $g_{tt}$ is constant, and (2) that in deriving its coefficients,
one never considers whether the selected components of the energy-momentum
tensor $T^{\mu\nu}$ are consistent with the assumption of homogeneity and
isotropy. In general relativity (GR), a constant lapse function arises only in
a free-falling frame, where the observer is not subject to any gravitational
influences. So even before we discuss the formal relationship between $g_{tt}$
and $T^{\mu\nu}$, proponents of the use of Equation~(1) to describe the
spacetime of an accelerating Universe need to explain why this basic tenet
of relativity theory should be violated, i.e., why in the case of FRW an
observer in the free-falling frame nonetheless sees a gravitational
acceleration without a corresponding time dilation.

The general class of spherically-symmetric spacetime metrics, of which
FRW is a special case, may be represented as
\begin{equation}
ds^2=e^{2\Phi/c^2}c^2dt^2-e^\lambda dr^2-R^2d\Omega^2\;,
\end{equation}
where $d\Omega^2\equiv d\theta^2+\sin^2\theta\,d\phi^2$, and $\Phi$,
$\lambda$, and $R$ are each functions of $r$ and $t$, and are to be
determined by solving Einstein's equations. It is somewhat tedious, though
straightforward, to derive the dependence of $g_{tt}\equiv e^{2\Phi/c^2}$
on $T^{tt}$, $T^{rr}$, $T^{\theta\theta}$ and $T^{\phi\phi}$. In the
conventional approach, however, $\Phi$ is set equal to $0$ to arrive at
Equation~(1) without following this procedure. As such, the FRW form
of the metric does not produce any time dilation relative to the
proper time in a local inertial frame, even in cases where the observer
sees an accelerated expansion of the spatial coordinates, i.e., when
$\ddot{a}\not=0$. This is critical because, as we shall see shortly,
the Hubble flow  in FRW is not inertial when $\ddot{a}\not=0$.

But as shown in ref.~[8], $g_{tt}$ is in fact not equal to one when
$\rho+3p\not=0$, a requirement of $\ddot{a}\not=0$. Borrowing a result
from that paper, we find that
\begin{equation}
e^{2\Phi(t)/c^2}=h{\dot{a}}^2e^{{\cal I}(t)}\;,
\end{equation}
where
\begin{equation}
{\cal I}(t)\equiv \int^t_0 dt^\prime\;{8\pi G\over 3c^2H}e^{\Phi/c^2}(\rho+3p)\;,
\end{equation}
and $H\equiv e^{-\Phi/c^2}(\dot{a}/a)$ is the Hubble constant. This expression
provides the necessary, formal relationship between the lapse function $g_{tt}$
and the active mass $\rho+3p$ as seen by an observer in the comoving frame.
Clearly, $\Phi$ changes for different values of the active mass and corresponding
expansion history encoded into the integral for ${\cal I}$ over cosmic time.
In order to achieve a constant $\Phi$, we must have ${\cal I}\rightarrow 0$,
which is guaranteed only when $\rho+3p\rightarrow 0$. Then one can show that
$\dot{a}$ is also constant, and we may set $g_{tt}=1$ with an appropriate
choice of the initial condition $h$.

This is where Kim, Lasenby and Hobson \cite{Kim2016} interject with a counter claim that
one is free to {\it choose} $\Phi=0$, in spite of its evident dependence on
$\rho+3p$. They do so without explaining why they view Equation~(3) as allowing
$\Phi$ to have different values for the same $\rho+3p$, and without providing
a physical justification for making their particular choice leading to $g_{tt}=1$.
In this regard, their approach is no different from the conventional procedure
of simply forcing $g_{tt}=1$ in Equation~(1) before the metric is introduced into
Einstein's equations, thereby removing any possibility of a time dilation. The
answer, of course, is that for the same $\rho+3p$, different values of $\Phi$
correspond to different coordinate systems. And in order to force $\Phi=0$,
while keeping ${\cal I}(t)\not=0$, they must consider the observer to be
{\it simultaneously} in the comoving frame where an active mass $\rho+3p\not=0$
produces acceleration (according to the second Friedmann equation,
in which $\ddot{a}\sim \rho+3p$) and $\Phi\not=0$, and in the free-falling
frame where $g_{tt}=1$ by choice. If the introduction
of the spherically-symmetric form of the metric (Equation~2) into
Einstein's equations were to reduce to the FRW spacetime shown in
Equation~(1) irrespective of how one chooses the active mass, then
this should happen automatically and unambiguously, {\it without} the manual
intervention by Kim et al. \cite{Kim2016} to force $\Phi=0$ in Equation~(3).
As we shall see below, this automatic reduction to Equation~(1)
does not happen because the comoving frame is not inertial, except
in the special case when $\rho+3p=0$.

This is also the reason why a gauge transformation cannot be used to
eliminate the coordinate-dependent $g_{tt}$ after the fact. Often
misunderstood and incorrectly used, a gauge transformation in GR
is a transformation of the coordinates selected in order to alter
the metric coefficients $g_{\mu\nu}$ to a desired form. Since
$\Phi$ is solely a function of $t$, it is straightforward to
find the necessary transformation for FRW but, as discussed
in ref.~[8], such a transformation would require the
existence of two distinct frames of reference: one (the
comoving) frame in which the observer sees $\ddot{a}\not=0$
when $\rho+3p\not=0$, and therefore $g_{tt}\not=1$, and a second
(free-falling) frame in which the observer sees no time dilation
and no acceleration. In other words, an observer cannot carry out
an actual gauge transformation and stay within the same frame. And since
the comoving frame in FRW is generally not inertial, an observer
who experiences gravitational effects that lead to a
coordinate-dependent lapse function cannot simultaneously
also be in a free-falling frame where he does not measure
a time dilation.

Ultimately, the issue we have been discussing here, and in
refs.~[8,9], has to do with whether or
not one can always clearly distinguish between accelerated and
inertial frames. The answer is yes (see, e.g., ref.~[10],
\S~3.2). In general relativity, acceleration can always be
measured absolutely, unlike velocity, which is only measurable
in a relative sense. For this reason, we can always find a local
diffeomorphism that reduces the chosen manifold's metric to a
Minkowski metric in a sufficiently small neighborhood of a given
spacetime point when tidal forces are 
ignored \cite{Harvey1964,Carrera2010}. The FRW metric is not asymptotically
flat and has a non-vanishing spacetime curvature tensor
$R_{\alpha\beta\gamma\delta}$. Thus, in spite of the fact
that its Weyl tensor is zero, which allows Equation~(1) to
be written as a conformally-flat metric, the comoving frame
used to write the FRW metric is not in free fall.

One seldom sees this property of FRW invoked in the interpretation
of measurements made by a Hubble observer 
(however, see, refs.~[13,14,15]), 
but one can easily demonstrate that the Hubble
flow in FRW is in fact not inertial when $\rho+3p\not=0$. The metric
coefficients $g_{\mu\nu}(x)$ and affine connections
$\Gamma^\lambda_{\;\,\mu\nu}(x)$ at $x^\alpha=(ct,r,\theta,\phi)$
contain enough information for us to determine the local inertial
coordinates $\xi^\alpha(x)$ in the neighborhood of $x^\alpha$.
As shown, e.g., in ref.~[10] or [13,14],
the local inertial coordinates $\xi^\alpha(x)$ satisfy the equation
\begin{equation}
{\partial^2\xi^\mu\over \partial x^\lambda\,\partial x^\kappa}=
\Gamma^\nu_{\;\,\lambda\kappa}(x){\partial\xi^\mu\over\partial
x^\nu}\;.
\end{equation}
In their most general form, the non-vanishing affine
connections in FRW are
\[ \hskip-0.7in \begin{array}{lll}
\Gamma^0_{\;\,00}={\dot{\Phi}/c^3} && \\
\Gamma^0_{\;\,rr}= {1\over c}a\dot{a}\,e^{-2\Phi/c^2} & \Gamma^0_{\;\,\theta\theta}= {1\over c} a\dot{a}r^2\,
e^{-2\Phi/c^2} & \Gamma^0_{\;\,\phi\phi}= {1\over c} a\dot{a}r^2\sin^2\theta\, e^{-2\Phi/c^2}  \\
\Gamma^r_{\;\,r0}= {1\over c}{\dot{a}\over a} & \Gamma^r_{\;\,\theta\theta}= -r & \Gamma^r_{\;\,\phi\phi}= -r\sin^2\theta   \\ 
\Gamma^\theta_{\;\,\theta 0}= {1\over c}{\dot{a}\over a} & \Gamma^\theta_{\;\,\theta r}= {1\over r} & 
\Gamma^\theta_{\;\,\phi\phi}= -\cos\theta\,\sin\theta \\
\Gamma^\phi_{\;\,\phi 0}= {1\over c}{\dot{a}\over a} & \Gamma^\phi_{\;\,\phi r}= {1\over r} & \Gamma^\phi_{\;\,\phi\theta}= \cot\theta \;.
\end{array} \]
Introducing these into Equation~(5), we see that the
inertial coordinates $\xi^\mu=(c\tilde{t},\tilde{r},\tilde{\theta},
\tilde{\phi})$ must satisfy the following expressions:
\begin{equation}
{\partial^2 \tilde{t}\over\partial t^2}={\dot{\Phi}\over c^2}{\partial
\tilde{t}\over t}\;,
\end{equation}
\begin{equation}
{\partial^2\tilde{t}\over \partial t\,\partial r}={\dot{a}\over a}{\partial
\tilde{t}\over \partial r}\;,
\end{equation}
\begin{equation}
{\partial^2\tilde{t}\over\partial r^2}={1\over c^2}\dot{a}a\,e^{-2\Phi/c^2}\,
{\partial\tilde{t}\over\partial t}\;,
\end{equation}
\begin{equation}
{\partial^2\tilde{r}\over\partial t^2}={\dot{\Phi}\over c^2}\,{\partial
\tilde{r}\over\partial t}\;,
\end{equation}
and
\begin{equation}
{\partial^2\tilde{r}\over\partial r\partial t}={1\over c}{\dot{a}\over a}\,
{\partial\tilde{r}\over \partial r}\;.
\end{equation}

Let us first consider the local inertial frame for an FRW
cosmology with zero active mass (i.e., $\rho+3p=0$), corresponding
to $a(t)=t/t_0$ (normalized such that $a=1$ today). We will place
the observer near the origin of his coordinates, so that $r/ct\ll 1$.
For this equation of state, we also have $\Phi=0$ (see Equation~3).
The solution to Equations~(6-10) is
\begin{equation}
\tilde{r}=a(t)r\;,
\end{equation}
and
\begin{equation}
\tilde{t}\approx t\left(1+{1\over 2}\left[{\tilde{r}\over ct}\right]^2\right)\;,
\end{equation}
correct to second order in $r/ct$. Thus, according to these expressions,
the local inertial frame in the vicinity of $r=0$ coincides with the
Hubble flow and, in this situation, one has $\ddot{a}=0$ with
$g_{tt}=1$. Thus, for this case ({\it and this case only}), it
is legitimate to consider the comoving and free-falling frames to
be identical, and for the observer to see no acceleration and no
time dilation, so that $\Phi=0$.

When $\rho+3p\not=0$, however, $a(t)$ is no longer linear in $t$,
and $\tilde{r}$ in Equation~(11) no longer satisfies Equations~(9)
and (10), regardless of whether or not one uses $\Phi=0$. For example,
in order for $\tilde{r}=ar$ to be consistent with Equation~(9), we
would need either $\ddot{a}=0$ (if $\Phi=0$), or $\ddot{a}-\dot{a}
\dot{\Phi}/c=0$ (if $\Phi\not=0$), both of which occur only when \cite{Melia2016}
$\rho+3p=0$. Of course, one can still find a local
inertial frame when $\rho+3p\not=0$, but it is not coincident with
the Hubble flow \cite{Harvey1964,Liu1987a,Liu1987b,Carrera2010,Kopeikin2015}.
In this instance, the solution
may be written as a polynomial to second order in $r$, as
demonstrated in \S~3.2 of ref.~[10], or [13,14],
but we don't need to reproduce its coefficients here. The fact that
the Hubble flow in FRW is not an inertial frame when $\rho+3p\not=0$
is sufficient to prove our argument. Therefore, setting $\Phi=0$
when $\rho+3p\not=0$ in the comoving frame is not consistent with
the fact that a time dilation should be measurable as a result of
the acceleration seen by the observer.

The confusion about the measurability of a time dilation due to acceleration
in FRW may be due to the fact that, because of homogeneity, $g_{tt}$ can
only be a function of $t$, not $r$, unlike the situation in the Schwarzschild
spacetime, where it is a function of $r$ and not of $t$. But this difference
in coordinates does not change the requirement that an acceleration must
always produce a time dilation that is measurable relative to the passage of
proper time in an inertial frame. Therefore, regardless of whether $g_{tt}$
is a function of $t$ or $r$, the time dilation resulting from an accelerated
expansion of the spatial coordinates cannot be ``hidden" from the observer.
Putting $g_{tt}=1$ in Equation~(1), even when $\ddot{a}\not=0$, suggests
otherwise, and is therefore inconsistent with basic relativity theory. To
guarantee that $\Phi=0$ in Equation~(3), one must have $\rho+3p=0$.

Concerned by ``gauge ambiguities'' in their first attempt at challenging
the zero active mass condition, Kim et al. \cite{Kim2016b} extended their
argument by attempting to derive the FRW solution using a tetrad, which 
they claim constitutes a superior method that avoids such gauge ambiguities. 
They apparently believe that the physics of a problem may be changed
merely by altering the calculational technique. But this is clearly false;
if a gauge amibuity is present with one approach, it is present for all 
approaches because, as we have emphasized all along, a gauge 
transformation is not arbitrary---it constitutes a transformation of 
coordinates from one frame to another. Kim et al. believe that the
sophistication of the tetrad approach somehow selects a unique 
frame of reference, creating their sought-after merger of a 
non-inertial comoving frame with a free-falling frame. This
thinking, however, is flawed. The tetrad and its inverse simply
package the metric coefficients and, once a solution is found,
they must be unpacked to restore the spacetime coordinates.
One therefore gets out what one puts in. In this regard, their
approach is similar to that of Tupper \cite{Tupper1974}, who
explicitly set $g_{tt}=1$ in his use of the tetrad field equations
to derive a generalized Friedmann equation. Kim et al.'s use of
the tetrad is more subtle, but they too invoked a special choice
of coordinates---specifically, a particular choice of time---in 
order to simplify their tetrad ansatz {\it before} using the
field equations to derive a solution. 

Finally, we briefly comment on Kim et al.'s attempt to discredit
the role of $R_{\rm h}$ as a true gravitational horizon. In so doing,
they follow the lead of van Oirschot et al. \cite{van2010} and
Lewis et al. \cite{Lewis2012}, who failed to realize that a horizon
is observer dependent. An observer ``sees'' a horizon only in terms
of null geodesics that actually reach him. Thus, to correctly interpret
the role played by the surface at $R_{\rm h}$, one must actually
solve the null geodesic equations, as demonstrated by Bikwa et
al. \cite{Bikwa2012} and Melia \cite{Melia2012}, not simply rely
on how or when $R_{\rm h}$ changes with time. A complete
formal discussion of how $R_{\rm h}$ delimits the portion of the
Universe visible to us today is given in Melia \cite{Melia2013}. 

\section{Concluding Remarks}
In this paper we have shown that the counter argument raised by Kim,
Lasenby and Hobson \cite{Kim2016} is simply based on the conventional approach
of forcing the lapse function of the metric to be constant, without
providing a physical justification for this particular ``choice," even
when the formal relationship between $g_{tt}=e^{2\Phi/c^2}$ and $T^{\mu\nu}$
permits a range of possibilities for the same active mass $\rho+3p$.
In reality, a specific selection of $\Phi$ corresponds to a particular
choice of coordinates. In FRW, the Hubble flow is non-inertial when
$\rho+3p\not=0$, so the comoving frame cannot be identical to
the free-falling frame. Therefore one cannot measure an acceleration
without a corresponding time dilation.

\noindent{\bf Acknowledgements:} I am grateful to PMO in Nanjing, China, for its hospitality while this work
was being carried out. This work was partially supported by grant 2012T1J0011 from The
Chinese Academy of Sciences Visiting Professorships for Senior International Scientists,
and grant GDJ20120491013 from the Chinese State Administration of Foreign Experts Affairs.

\end{document}